# A Computational Study of Negative Surface Discharges: Characteristics of Surface Streamers and Surface Charges

**Xiaoran Li** and **Anbang Sun**
State Key Laboratory of Electrical Insulation and Power Equipment, School of Electrical Engineering,
Xi'an Jiaotong University, Xi'an, 710049, China

**Jannis Teunissen**
Centrum Wiskunde & Informatica,
Amsterdam, The Netherlands

## ABSTRACT

We investigate the dynamics of negative surface discharges in air through numerical simulations with a 2D fluid model. A geometry consisting of a flat dielectric embedded between parallel-plate electrodes is used. Compared to negative streamers in bulk gas, negative surface streamers are observed to have a higher electron density, a higher electric field and higher propagation velocity. On the other hand, their maximum electric field and velocity are lower than for positive surface streamers. In our simulations, negative surface streamers are slower for larger relative permittivity. Negative charge accumulates on a dielectric surface when a negative streamer propagates along it, which can lead to a high electric field inside the dielectric. If we initially put negative surface charge on the dielectric, the growth of negative surface discharges is delayed or inhibited. Positive surface charge has the opposite effect.

## 1 INTRODUCTION

**SURFACE** discharges are common in electronics and high voltage devices. Dielectrics not only distort nearby electric fields but also serve as a possible electron sink or source. Dielectrics can therefore play a critical role in the formation and propagation of discharges [1]. We have recently explored the interaction between positive streamers and dielectrics in [2]. In this paper, we investigate the properties of negative surface discharges, which can have quite different characteristics from positive ones [3, 4].

In the last decades, experimental studies of negative surface discharges have often focused on the measurement of flashover voltages [5] and surface charge accumulation [6]. There have also been several studies on the effect of surface charge on the subsequent breakdown [7–11]. Surface dielectric barrier discharges (SDBDs) have also been studied experimentally. In e.g. [12], advanced diagnostic were used to measure streamer velocities and electric fields. Such experimental studies can provide practical guidelines for insulation engineers. However, performing a microscopic investigation on the plasma-surface interaction, especially at atmospheric pressure, is extremely challenging, as a non-intrusive diagnostic method with a spatial resolution down to micrometers and a temporal resolution down to nanoseconds is required [13]. To gain further insight into negative surface discharges, numerical simulations have also been performed. We highlight a few examples below.

Tran *et al.* [14] performed 2D axisymmetric simulations of negative corona and barrier discharges in a needle-to-plane geometry. They validated the model parameters by comparing with experimental data. Sima *et al* [15] used a 2D axisymmetric fluid model to identify different surface discharge stages from the electric current, in a geometry consisting of two plate electrodes and a cylindrical insulator. The resulting surface charge and the effects of the voltage amplitude and the dielectric properties were also investigated. Numerical 2D simulations of nanosecond-pulsed SDBDs of positive and negative polarity have also been performed. In [16] and [3], the near-surface discharge structure and electric field were analyzed, with the latter also focusing on secondary electron emission.

In past research on surface discharges, many different geometries have been considered. Here, we consider a geometry in which a flat dielectric is placed between parallel-plate electrodes, as in our previous work on positive streamers [2]. Such a geometry is relevant for applications in HV insulation. We simulate negative streamers interacting with dielectrics, including discharge inception, attachment to the dielectric and propagation over the surface. We also study the effect of the applied voltage, the relative permittivity and preset surface charge on negative surface discharges.

The paper is organized as follows. The simulation model is described in Section 2. In Section 3.1, we focus on the interaction between negative streamers and dielectrics and on surface charge accumulation during streamer propagation. Then the effects of the applied voltage (Section 3.2) and the relative permittivity (Section 3.3) are investigated. Finally, the effect of pre-set surface charge on negative surface discharges is studied in Section 3.4.

## 2 SIMULATION MODEL

We use the same simulation setup and model as for our study of positive streamers [2], so that results can directly be compared. The simulation model and setup are briefly introduced below, for further details we refer to [2].

### 2.1 SIMULATION SETUP

The geometry we use consists of a flat dielectric placed between two parallel-plate electrodes, as shown in Figure 1. This geometry resembles some actual HV insulation applications, and its simplicity makes it suitable for numerically studying surface discharges. The computational domain measures (40 mm)$^2$, and the dielectric is placed on the left side with a width of 10 mm. Direct high voltage is applied at the upper electrode, and the lower electrode is grounded. The gas is artificial air (80% $N_2$ and 20% $O_2$) at 1 bar and 300 K. The background densities of electrons and positive ions are set to $10^{10}$ m$^{-3}$. Discharges usually start in regions where the electric field is locally enhanced. In actual HV devices, the electric field is often enhanced at a triple junction between gas, dielectric and electrode. A realistic description of discharge inception (due to e.g. partial discharges and surface charge accumulation) is outside the scope of the present paper. Instead, an ionized seed is placed near the upper triple junction to enhance the electric field locally, as indicated in Figure 1. The seed we used here is about 2 mm long with a radius of about 0.4 mm. Its top edge just touches the upper electrode. Initially, the seed is electrically neutral, with electron and positive ion densities of $5 \times 10^{18}$ m$^{-3}$ at the center, decaying smoothly from a radius of 0.2 mm to zero at 0.4 mm, see [2] for details.

The distance $d$ between the initial seed and the dielectric is slightly varied in the paper, see Table 1. In Section 3.1, we use $d = 1$ mm to study the attraction of streamers towards the dielectric. In Section 3.2 and 3.3, we use $d = 0.5$ mm, and in Section 3.4 we use $d = 0$ mm so that discharges directly start at the interface. When the initial seed is placed farther away from the dielectric, it will take longer for the streamer to reach the dielectric, but the further discharge evolution is similar, as was also observed in [2].

The applied voltage, relative permittivity and pre-set surface charge are varied in Sections 3.2, 3.3 and 3.4, see Table 1. We study how these parameters affect negative streamers, in particular their inception, propagation, morphology and surface charge characteristics. The simulations are performed up to 20 ns. In all considered cases, streamers have reached the dielectric and propagated over it within 20 ns. We do not consider later stages, in which the discharge has reached the other electrode.

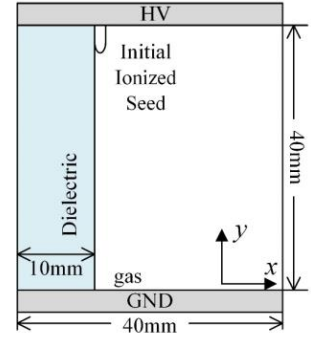

**Figure 1.** Schematic of the computational domain. Unless indicated otherwise (see Table 1), -120 kV is applied at the HV electrode and the relative permittivity of the dielectric is $\varepsilon_r = 2$.

**Table 1.** Investigated parameters and their values in each section*.

| Section | $d$ (mm) | $U$ (kV) | $\varepsilon_r$ | $\sigma_s$ (pC/mm$^2$) |
|---|---|---|---|---|
| 3.1 | 1 | -120 | 2 | 0 |
| 3.2 | 0.5 | (-112, -120, -128) | 2 | 0 |
| 3.2 | 0.5 | -120 | (2, 3, 5) | 0 |
| 3.4 | 0 | -120 | 2 | (-5, -1, 0, 1, 5) |

*: Here $d$ is the distance between seed center and the dielectric surface; $U$ the applied voltage; $\varepsilon_r$ the relative permittivity of the dielectric and $\sigma_s$ the initial surface charge.

### 2.2 PLASMA MODEL

A 2D fluid model is used in this paper, which is based on Afivo-streamer [17] and improved to include dielectric surfaces [2]. It uses the adaptive mesh refinement and the parallel multigrid solver provided by the underlying Afivo framework [18].

The fluid model used here is of the drift-diffusion reaction type with the local field approximation. The model keeps track of the electron density $n_e$, the positive ion density $n_i^+$ and the negative ion density $n_i^-$, which evolve in time as:

$$\frac{\partial n_e}{\partial t} = -\nabla \cdot (-n_e \mu_e \mathbf{E} - D_e \nabla n_e) + S_i - S_a + S_{pi}$$
$$\frac{\partial n_i^+}{\partial t} = -\nabla \cdot (n_i^+ \mu_i^+ \mathbf{E}) + S_i + S_{pi} \quad (1)$$
$$\frac{\partial n_i^-}{\partial t} = -\nabla \cdot (-n_i^- \mu_i^- \mathbf{E}) + S_a$$

Here, $\mu_e$ is the electron mobility, $D_e$ the electron diffusion coefficient, $\mathbf{E}$ the electric field, and $\mu_i^{+/-}$ the positive/negative ion mobilities. We use $\mu_i^+ = 3 \times 10^{-4}$ m$^2$/Vs and $\mu_i^- = 0$, consistent with [2]. The electron impact ionization and electron attachment terms are given by $S_i = \alpha \mu_e E n_e$ and $S_a = \eta \mu_e E n_e$, respectively, where $\alpha$ and $\eta$ are the ionization and attachment coefficients. The production of photoelectrons from photoionization is included with the term $S_{pi}$.

We use a Monte Carlo approach to implement Zheleznyak's photoionization model, in which discrete ionizing photons are generated and absorbed using random numbers. Their absorption at the dielectric is taken into account. The photoionization source term is updated every 10 time steps using $10^5$ 'virtual' photons. A detailed description of the photoionization procedure can be found in [2] and in [19].

The local field approximation is used, so that $\mu_e$, $D_e$, $\alpha$ and $\eta$ are functions of the local electric field strength. Electron transport and reaction coefficients for air (1 bar, 300 K) were generated with Monte Carlo particle swarm simulations, using Phelps' cross sections [20]. In this work, these coefficients are tabulated up to a certain maximum electric field, which is here 35 kV/mm; for higher fields, we use the tabulated value at 35 kV/mm.

Electrons and ions attach to the dielectric surface when they flow onto it. They then locally contribute to the surface charge density $\sigma_s$ at the dielectric-gas interface. Reactions or diffusion on the surface are not taken into account, so $\sigma_s$ changes in time as:

$$\partial_t \sigma_s = -e(\Gamma_e + \Gamma_i^-) + e\Gamma_i^+ \quad (2)$$

Here $e$ is the elementary charge and the other terms correspond to the fluxes towards the gas-dielectric interface: $\Gamma_e$ for electrons, $\Gamma_i^-$ for negative ions and $\Gamma_i^+$ for positive ions. We calculate fluxes on the gas-dielectric interface in the same way as fluxes in the bulk gas, which may not always be accurate [21].

However, we expect that this approximation, which was also used in e.g. [1, 22, 15], has no strong effect for the transient (non-equilibrium) simulations presented here. The minimum grid spacing $\Delta x$ used for the adaptive mesh is about 1.2 $\mu$m. The mesh refinement depends on the local ionization coefficient $\alpha$, ensuring that $\Delta x < 1/\alpha$.

The electric field **E** is calculated by solving Poisson's equation:
$$\nabla \cdot (\varepsilon \nabla \varphi) = -(\rho + \delta_s \sigma_s) \quad (3)$$
$$\mathbf{E} = -\nabla \varphi$$

where $\varepsilon$ is the dielectric permittivity, $\rho$ is the volume charge density, and $\delta_s$ maps the surface charge $\sigma_s$ on the gas-dielectric interface to the grid cells adjacent to the dielectric. At the interface, the normal component of the electric field satisfies the classic jump condition:

$$\varepsilon_1 E_1 - \varepsilon_2 E_2 = \sigma_s \quad (4)$$

where $\varepsilon_1$ and $\varepsilon_2$ represent the permittivities on both sides of the interface, and $E_1$ and $E_2$ are the electric field components normal to the interface.

For positive streamers, secondary electron emission (SEE) from a dielectric can be important, because these electrons can start avalanches growing towards the streamer head. For negative streamers, electrons move away from the streamer head, so that SEE electrons released from the dielectric would immediately flow back onto it. SEE from dielectrics is therefore neglected in this paper. We remark that SEE could play a role in the initiation of negative streamers (for example through surface charge accumulation), but that is outside the scope of the present paper.

## 3 RESULTS & DISCUSSION

### 3.1 INTERACTION BETWEEN NEGATIVE STREAMERS AND DIELECTRICS

#### 3.1.1 Comparison with positive streamers

The attraction of positive streamers to dielectrics has been demonstrated in several experiments (e.g. [23]) and simulations (e.g. [2]). In our simulations, we observe a similar attraction for negative streamers. Figure 2a shows the development of a negative streamer between 4 ns and 14 ns for an initial seed placed 1 mm away from the dielectric surface. For comparison, the development of a positive streamer under the same conditions (but with a different voltage polarity) is shown in Figure 2b.

The electron density in the positive streamer channel (~$10^{19}$ m$^{-3}$) is higher than in the negative channel (~$10^{18}$ m$^{-3}$). This can be explained as follows. Electrons drift away from negative streamers, whereas they drift towards positive streamers. The charge layer around positive streamers is therefore formed by positive ions, which are less mobile than electrons, so that positive streamer channels are more concentrated [24]. However, for both polarities, the electron densities of surface streamers (~$10^{21}$ m$^{-3}$) are higher than those of gas streamers, which we also observed in [2]. This is primarily due to the enhanced electric field of surface streamers, shown in Figure 3 and discussed below. Surface streamers have a higher field due to electrostatic effects and due to their reduced radius compared to gas streamers.

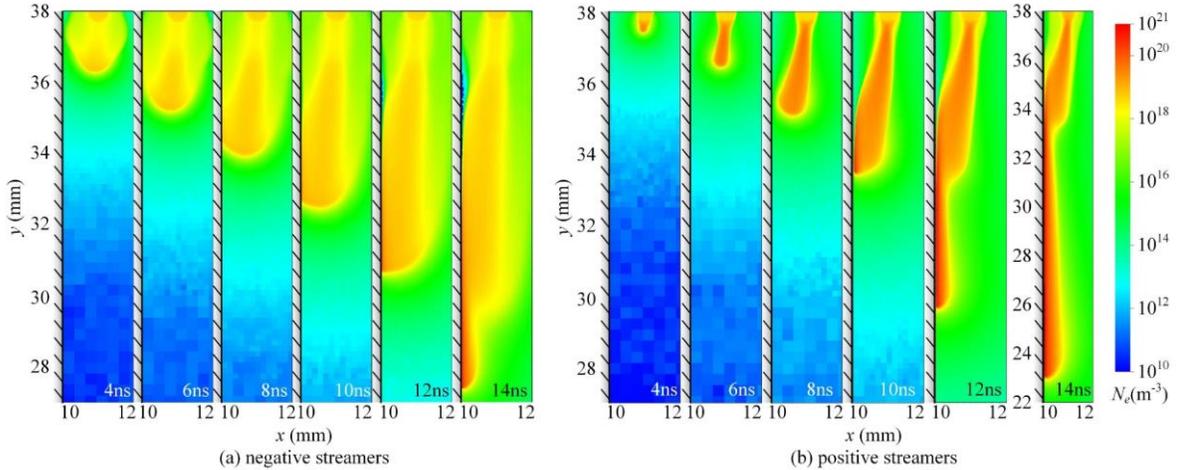

**Figure 2.** Evolution of negative (a) and positive (b) streamers between 4 ns and 14 ns, for an initial seed located at 1 mm from the dielectric surface on the left. The applied voltage is -120 kV for negative streamers and 120 kV for positive streamers. The relative permittivity is 2. Note that only part of the computational domain is shown in this figure.

Another distinguishing feature is that the negative streamer starts earlier. At 4 ns, its length is about 2 mm, whereas the positive streamer just starts. However, afterwards positive streamers have a higher velocity, especially when propagating over the surface. This is illustrated in Figure 3, which shows the streamer velocity and its maximal electric field versus streamer length.

From Figures 2 and 3, we find that both negative and positive streamers reach the dielectric at around y = 35 mm. The negative surface streamer forms at around y = 29 mm and the positive surface streamer forms at about y = 33 mm. For both polarities, the maximum electric field and streamer velocity increase when propagating over the surface. The maximum electric field for the negative surface streamer is about 20 ~ 25 kV/mm; for the positive one, it is over 30 kV/mm.

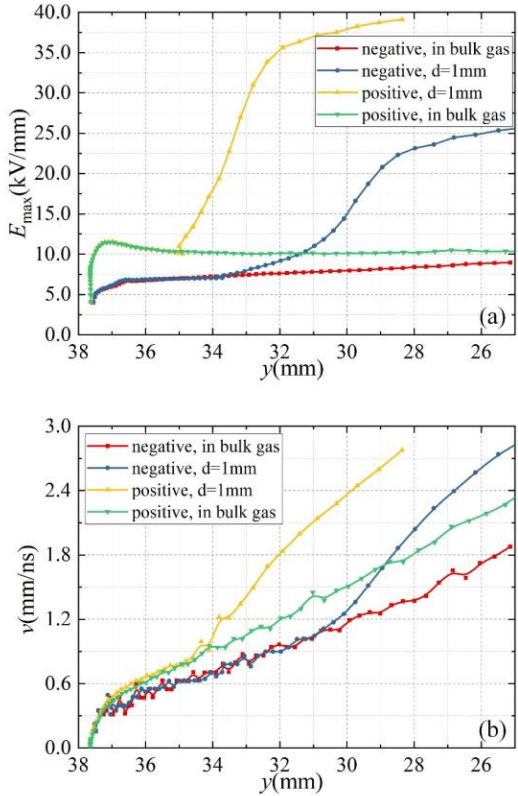

**Figure 3.** Streamer maximal electric field (a) and velocity (b) versus y-location of the electric field maximum. Results are shown for the negative and positive streamers in Figure 2 (labeled "negative, d=1mm" and "positive, d=1mm", respectively) and for corresponding cases in bulk gas without a dielectric. The streamer velocity $v$ is calculated by dividing the distance the streamer head moves between two consecutive outputs by the output time interval.

For both polarities, the dielectric's polarization strengthens the electric field between the streamer and the dielectric, which attracts the streamer to the dielectric. However, the negative streamer propagates along the surface for 6 mm before a surface streamer forms, whereas this distance is only 2 mm for the positive streamer. There can be two reasons for this. First, for negative streamers, electrons move away from the streamer channel, which leads to the accumulation of negative surface charge on the dielectric (see Section 3.1.2 for more details). This surface charge lowers the electric field between the streamer and the dielectric. Second, the negative streamer has a larger radius and a lower electric field. This means it has lower and more spread out charge density at its head, which leads to weaker electrostatic attraction to the surface.

Figure 4 shows the streamer velocity versus maximum electric field for the positive and negative streamers in Figure 2. Compared to streamers in bulk gas [4, 24], the relation between $v$ and $E_{max}$ is more complicated for streamers interacting with dielectrics. Three stages with different slopes can be distinguished. When $v < 0.9$ mm/ns, streamers are propagating towards the dielectric. For $v$ between 0.9 mm/ns and 1.6 mm/ns, a surface streamer forms, and for $v > 1.6$ mm/ns a surface streamer is propagating over the dielectric. Note that for the same velocity, negative streamers have a lower maximum electric field, but that the three stages occur at similar streamer velocities for both polarities.

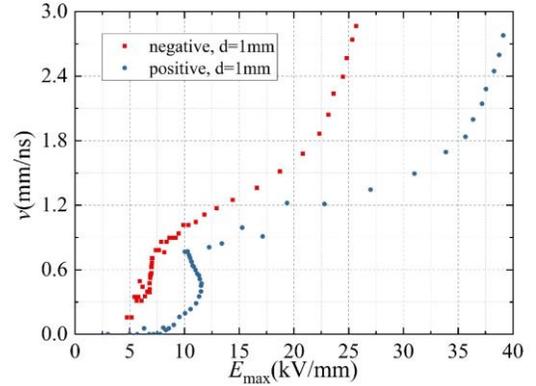

**Figure 4.** Streamer velocity versus maximum electric field at the streamer head. Results are shown for the negative and positive streamers in Figure 2.

### 3.1.2 Surface charge characteristics

As mentioned before, electrons from a negative surface discharge move outwards, so towards the dielectric it is propagating over. Figure 5 shows the evolution of the surface charge for the negative streamer shown in Figure 2a. Up to 12 ns, the surface charge only increases, which happens most rapidly near the streamer head. Afterwards, a reduction in surface charge behind the streamer head is visible. This happens when the back of the negative streamer becomes more positively charged, so that the field between the back of the streamer and the negatively charged surface reverses. Positive ions then flow to the surface and partially neutralize it.

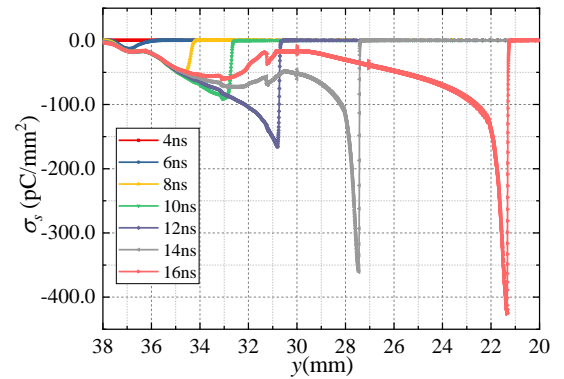

**Figure 5.** The evolution of the dielectric surface charge from 4 ns to 16 ns for the negative streamer from Figure 2a.

The increasing surface charge near the streamer head can produce a high electric field inside the dielectric, which was also observed in [3]. Figure 6 shows the electric field distribution for the streamer in Figure 2a at 14 ns. A high electric field is present around y = 37.43 mm, which corresponds to the location of the peak of the surface charge at 14 ns in Figure 5.

We remark that for positive surface streamers [2, 3], a streamer-dielectric gap with a high electric field but a low electron density has been observed. For negative surface streamers no such gap is present, and the streamers can fully connect to the dielectric surface, as shown in Figure 6.

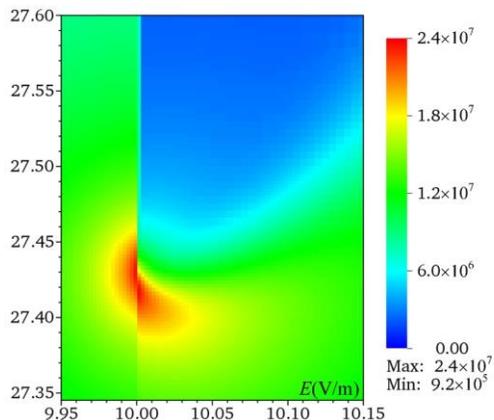

**Figure 6.** Electric field distribution for the negative surface streamer of Figure 2a at 14 ns.

## 3.2 EFFECT OF APPLIED VOLTAGE

To study the effect of the applied voltage on negative surface discharges, we have performed several simulations for applied voltages of 100 to 128 kV, which correspond to background electric fields of 2.5 to 3.2 kV/mm. In all cases, the initial seed was located at 0.5 mm from the dielectric, and the evolution up to 20 ns was simulated. Negative streamers usually require a higher background electric field than positive streamers [4]. With the geometry and initial seed used here, the formation of negative streamers required a background electric field of 2.6 kV/mm, which is a little bit lower than the breakdown threshold, whereas positive streamers could start in a field of 2.3 kV/mm [2]. We remark that with a different initial seed or with a pointed electrode streamers can also form in lower background fields.

Figure 7 shows electron densities for negative streamers in background electric fields of 2.8, 3.0 and 3.2 kV/mm. When compared at the same time, streamers are longer in a higher background electric field. Whereas the differences are initially small, they increase at later times, because the streamers accelerate. This is consistent with our findings for positive streamers [2]. Similar behavior was also observed experimentally, e.g. in [13]. Although the background electric field affects the streamer velocity, the overall development for these three cases is similar. Surface streamers form at about y = 32 mm, and when compared at the same length they have a similar shape.

The streamer velocity versus y-position of the streamer head is shown in Figure 8. With time, the velocities as well as the differences between them increase. Note that the negative streamer velocity does not start at zero, which is the case for positive streamers [2]. The difference is that negative streamers propagate with at least the electron drift velocity [24], whereas positive streamers can only grow due to ionization.

Figure 9 shows the surface charge distribution when the streamer heads are located close to y = 28 mm. The profiles are similar, so the background electric field has only a small effect on the amount of surface charge deposited at a certain length.

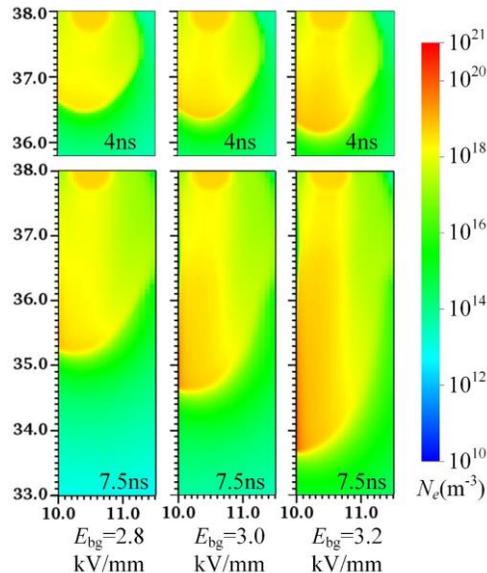

**Figure 7.** Electron densities for negative streamers in a background electric field of 2.8, 3.0 and 3.2 kV/mm, at 4 ns and 7.5 ns.

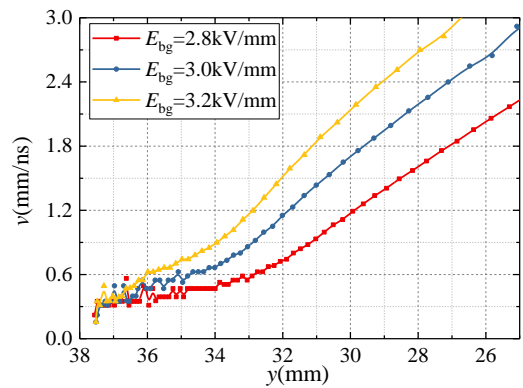

**Figure 8.** The streamer velocity versus the y-position of the streamer head in several background electric fields.

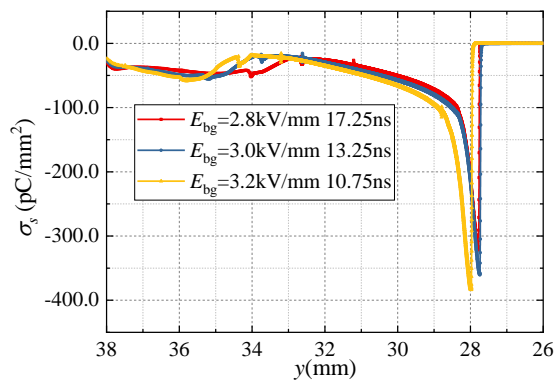

**Figure 9.** The dielectric surface charge for streamers in different background electric fields. Curves are shown at the moment the streamer heads are close to y = 28 mm.

## 3.3 EFFECT OF PERMITTIVITY

To study the effect of the dielectric permittivity on negative surface discharges, we have performed simulations with relative permittivities of 2, 3 and 5. The initial seeds were again located at 0.5 mm from the dielectric, and simulations ran up to 20 ns.

Figure 10 shows the electron density at 4 ns and 9 ns. At 4 ns, the streamer lengths are still similar to each other. However, at 9 ns, the streamer velocity is clearly higher with a lower relative permittivity. The same can be seen in Figure 11, which shows the streamer velocity versus the y-location of the streamer head. Initially, the streamer velocities are similar, but afterwards streamers are slower with a higher relative permittivity. The velocity difference (compared at the same length) becomes smaller as the streamers grow longer. The slower velocity can be explained from the following two aspects. A higher relative permittivity, which enhances the electric field between streamers and dielectrics, leads to stronger attraction of electrons to the surface. This directly leads to increased negative surface charge, which *reduces the electric field at the streamer head*. The other effect is that free electrons are more strongly attracted towards the dielectric. This can *reduce the amount of impact ionization taking place in front of the streamer*, as electron avalanches end up at the dielectric surface.

We remark that positive streamers behave differently: a larger relative permittivity led to faster discharge inception, but had almost no effect on the streamer velocity [2].

Figure 12 shows the surface charge distribution when the streamers are close to y = 30 and 28 mm. For streamers of the same length, there is more negative surface charge near the streamer head with a higher relative permittivity. After the streamer head has passed by, the surface charge profiles are similar for the three cases. We can deduce the amount of surface charge remaining after flashover is not sensitive to the dielectric permittivity. This is consistent with the discharge simulations reported in [15], in which the amount of surface charge was similar for different dielectric materials. On the other hand, the rate at which surface charge builds up before flashover could be sensitive to the permittivity.

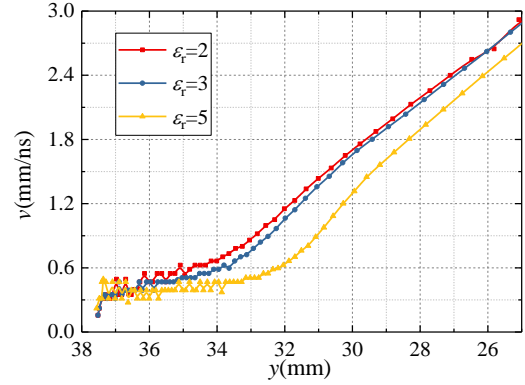

**Figure 11.** Streamer velocity versus y-location for different dielectric permittivities.

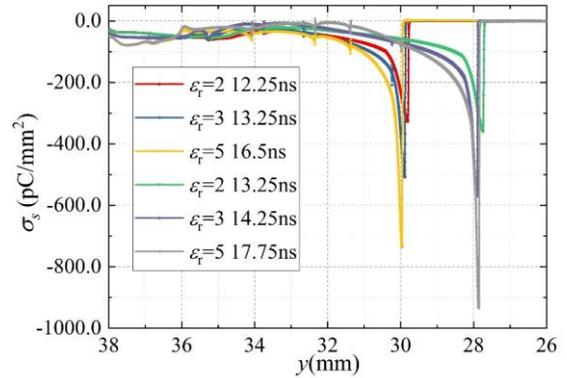

**Figure 12.** The dielectric surface charge for different relative permittivity, shown when the streamer heads are close to y = 28 and 30 mm. A higher relative permittivity leads to more negative charge close to the streamer head.

## 3.4 EFFECT OF PRESET SURFACE CHARGE

Surface charge accumulation is considered to be a tough problem for HVDC spacers [25]. There have been quite a few experimental studies on how surface charge affects subsequent discharges. Two cases can be considered: 'same-polarity' surface charge, which has the same polarity as the surface discharge, and 'opposite-polarity' surface charge. In two studies [8, 11], same-polarity surface charge increased flashover resistance, whereas surface opposite-polarity surface charge reduced flashover voltage levels. In contrast, another study found almost no effect of same-polarity surface charge [9], and in [10] both unipolar and mixed-polarity surface charge reduced flashover resistance. Therefore, the effect of preset surface charge on surface discharges remains inconclusive.

The different experimental results mentioned above could be caused by different charge deposition methods. The experimental surface charge deposition methods also create ionization (electrons and ions) in the gas. Since this ionization affects the formation of surface discharges [12], it is hard to single out the effect of the deposited surface charge. Differences could also be caused by fact that experimental charge deposition methods usually lead to a non-uniform charge distribution. A non-uniform surface charge distribution can enhance the electric field near some parts of the dielectric, while reducing it in others.

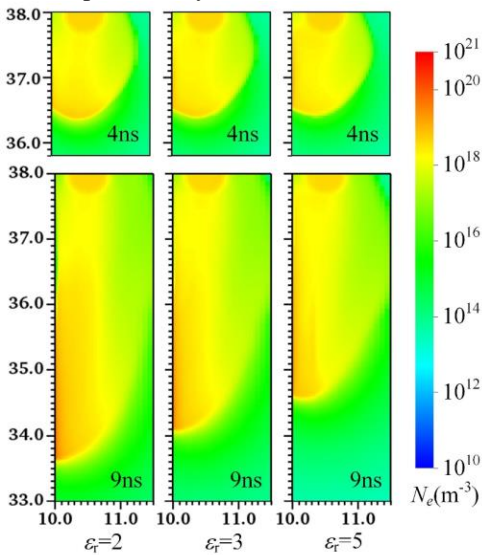

**Figure 10.** Streamer electron densities for dielectrics with relative permittivities $\varepsilon_r$ of 2, 3 and 5, shown at 4 and 9 ns.

Here we use numerical simulations to investigate the effect of preset surface charge on negative surface discharges. Compared to experiments, simulations allow full control over the initial surface charge distribution without affecting the background ionization level. Different amounts of surface charge (both positive and negative) are added at the beginning of the simulation. We place the initial seed so that its center coincides with the dielectric surfaces. This ensures that discharges start at the interface, which is also likely to happen in actual HV equipment. The simulations are performed up to 20 ns.

Figure 13 shows the maximum electric field versus time for preset surface charge densities of -5, -1, 0, 1 and 5 pC/mm$^2$. The surface charge is placed uniformly. An enhancement of the maximum electric field indicates the development of negative surface streamers, see Figure 3. Figure 13 therefore shows that preset positive surface charge accelerate the development of negative streamers around dielectrics, while negative surface charge delays or inhibits negative surface discharges. Negative surface charge reduces the electric field ahead the initial ionized seed, while positive charge enhances it. Our results agree with the experimental measurements in [8] and [11]. They are also in agreement with [13], in which it was found that residual surface charge with the same polarity as the applied high-voltage suppressed the development of surface discharges.

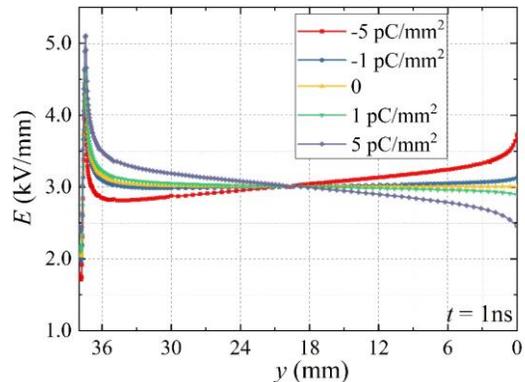

**Figure 14.** The initial electric field along the dielectric surface for simulations with preset surface charge densities of -5, -1, 0, 1 and 5 pC/mm$^2$. The values were measured at 1 $\mu$m outside the dielectric (in the gas) at $t = 1$ ns, when streamers start to form.

## 4 CONCLUSIONS

In this paper, the interaction between negative streamers and dielectrics has been studied with numerical simulations. We analyzed the main features of negative surface streamers and compared them to their positive counterparts. We studied how negative surface discharges are affected by the applied voltage, by the relative permittivity and by preset surface charge. Our main conclusions are:

1) Like positive streamers, negative streamers close to a dielectric will be attracted to it and form surface streamers. Compared to negative streamers in bulk gas, negative surface streamers have a higher maximum electric field, a higher electron density, and a higher propagation velocity. Compared to positive surface streamers, their maximum electric field and propagation velocity are slightly lower.

2) When negative surface streamers propagate along a dielectric, the dielectric becomes negatively charged. The peak charge density occurs around the streamer head, and it produces a high electric field inside the dielectric.

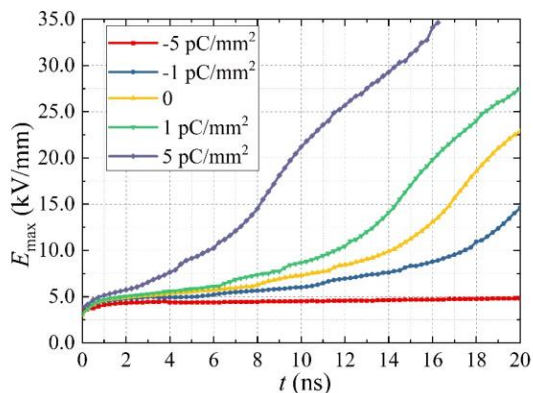

**Figure 13.** The maximum electric field versus time for simulations with preset surface charge densities of -5, -1, 0, 1 and 5 pC/mm$^2$. Enhanced electric fields indicate a negative surface streamer has formed.

Figure 14 shows the electric field along the dielectric surface for the above simulation cases, measured 1 $\mu$m away from the dielectric (in the gas) at t = 1 ns. At this time, the streamers start to form at the tip of the initial seed, located at 37.5 mm. Note that the uniform surface charge leads to a non-uniform change in the electric field, with the largest differences occurring near the electrodes. With a negative surface charge, the electric field near the negative HV electrode is reduced, whereas the field near the grounded electrode is enhanced. A positive surface charge has the opposite effect. These changes in the electric field have a strong effect on the development of surface discharges, as shown in figure 13.

In Section 3.1, we presented results in which negative streamers deposited negative surface charge on dielectrics. Such inhomogeneous surface charge may not increase discharge resistance. In actual devices, the effects of surface conduction and volume conduction should also be taken into account [25]. Further work is required to understand the role of these different mechanisms.

3) A higher applied voltage leads to a higher streamer velocity, but the accumulated surface charge at a given length is similar.

4) A higher relative permittivity slows down the development of negative surface streamers. This could be due to an increase in negative surface charge near the streamer head, which reduces the streamer's electric field. The effect becomes weaker for longer streamers.

5) Preset positive surface charge accelerates the development of negative streamers around dielectrics, whereas negative surface charge delays or inhibits negative surface discharges.

## ACKNOWLEDGMENT

This project was supported by the National Natural Science Foundation of China (51777164), the State Key Laboratory of Electrical Insulation and Power Equipment (EIPE18203), and the Fundamental Research Funds for the Central Universities of China (xtr042019009).

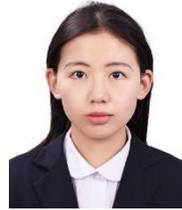

**Xiaoran Li** was born in Hebei, China, in 1996. She received the B.Sc. degrees from Xi'an Jiaotong University, Xi'an, China, in 2017. Currently, she is pursuing the Ph.D. degree in the School of Electrical engineering, Xi'an Jiaotong University. Her research interests are the discharge mechanism and simulation with dielectric surfaces.

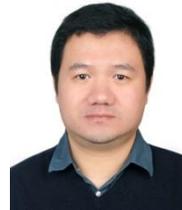

**Anbang Sun** was born in Anhui, China. He received the B.Sc degree and PhD degree from Northwestern Polytechnical University, in 2016 and 2010, respectively. From 2010 to 2011, he was a postdoc at ISAE, Toulouse, France. He was a postdoc researcher with Centrum Wiskunde & Informatica (CWI), Amsterdam, The Netherlands, from 2011 to 2014. He worked at Leibniz Institute for Plasma Science and Technology (INP Greifswald), Germany, as a scientist, from 2014 to 2016. He is currently a Professor with the school of electrical engineering, Xi'an Jiaotong University. His research interests include gas discharges, plasma propulsion technology, advance numerical skills for low temperature plasmas. He can be reached at anbang.sun@xjtu.edu.cn (corresponding author).

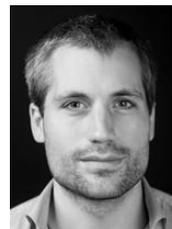

**Jannis Teunissen** was born in Amsterdam, The Netherlands. In 2015, he received a PhD degree from Eindhoven University of Technology for his work on the modeling of electric discharges, which was performed at Centrum Wiskunde & Informatica (CWI), Amsterdam. From 2016 to 2018, he was a postdoc at KU Leuven in Belgium. Since 2018, he is a research staff member at CWI. His interests are scientific computing, computational plasma physics and machine learning. He can be reached at jannis.teunissen@cwi.nl (corresponding author).